\documentclass[preprint,12pt]{elsarticle}
\usepackage[T1]{fontenc}
\usepackage[utf8]{inputenc}
\usepackage{amsfonts}
\usepackage{multirow,dcolumn}
\usepackage{booktabs}
\usepackage[version=3]{mhchem}
\usepackage{lmodern}
\usepackage{lipsum}
\usepackage{lineno}
\usepackage{threeparttable}
\usepackage[thinlines]{easytable}
\usepackage{caption}
\usepackage{tabularx}
\usepackage[dvipsnames]{xcolor}
\usepackage{blindtext}
\usepackage{booktabs}
\usepackage[version=3]{mhchem}
\usepackage[superscript,biblabel]{cite}
\usepackage{subfig}
\usepackage{graphicx}
\usepackage{adjustbox}
\usepackage{titlesec}
\usepackage{dcolumn}
\newcolumntype{d}[1]{D{.}{.}{#1}}

\usepackage{lmodern}
\usepackage{mathrsfs}
\usepackage{mathtools}
\usepackage{braket}

\newcommand{\sS}{\prescript{1}{}{\mbox{S}}}
\newcommand{\sP}{\prescript{1}{}{\mbox{P}}}
\newcommand{\sD}{\prescript{1}{}{\mbox{D}}}

\newcommand{\TT}{\mathcal{T}}

\newcommand{\SSS}{\mathcal{S}}

\newcommand{\tr}{(3)}
\newcommand{\pr}{(1)}
\newcommand{\dr}{(2)}
\newcommand{\dzr}{^{(0)}}
\newcommand{\dpr}{^{(1)}}
\newcommand{\ddr}{^{(2)}}
\newcommand{\dtr}{^{(3)}}
\newcommand{\bx}{\mathbf{x}}
\newcommand{\proj}{\hat{\mathscr{P}}}
\newcommand{\kp}{K^p}
\newcommand{\kr}{K^r}
\newcommand{\rqq}{R_q}
\newcommand{\rs}{R_s}
\newcommand{\resnd}[1]{\mathscr{\hat{R}}_{#1}}
  \newcommand{\resn}[1]{\mathscr{\hat{R}}_{{#1},\omega}}

  \newcommand{\resnzd}[1]{\hat{\mathscr{R}}_{{#1}}}

\newcommand{\op}[3]{{#1}_{#2}^{(#3)}}
\newcommand{\errname}{$\Delta_\mathrm{rel}$}
\newcommand{\relerror}{$|\delta \mu|/|\mu_\mathrm{exp}| \times 100 \% $}
\newcommand{\bb}[2]{\braket{ #1|#2}} 
\newcommand{\comute}[2]{ [ #1,#2]}
\newcommand{\cm}[2]{ [ #1,#2]}
\newcommand{\dg}[1]{#1^\dagger}
\newcommand{\dgg}{^\dagger}

\newcommand{\equ}[1]{\begin{equation} #1 \end{equation}}
\newcommand{\equl}[2]{\begin{equation}\label{#2} #1 \end{equation}}
\newcommand{\equm}[1]{\begin{multline} #1 \end{multline}}
\newcommand{\equml}[2]{\begin{multline}\label{#2} #1 \end{multline}}
\newcommand{\equa}[1]{\begin{align} #1 \end{align}}
\newcommand{\equal}[2]{\begin{align}\label{#2} #1 \end{align}}
\newcommand{\equs}[1]{\begin{equation}\begin{split} #1 \end{split}\end{equation}}
\newcommand{\equsl}[2]{\begin{equation}\begin{split}\label{#2} #1 \end{split}\end{equation}}

\newcommand{\fr}[1]{Eq.~(\ref{#1})}
\newcommand{\Frt}[1]{Table~\ref{#1}}
\newcommand{\Frf}[1]{Figure~\ref{#1}}
\newcommand{\frs}[1]{Section~\ref{#1}}
\newcommand{\rar}{\rightarrow}
\newcommand{\odps}[1]{\braket{\braket{X;Y}}_\omega^{[#1]}}
\newcommand{\odpr}[1]{\braket{\braket{X;Y}}_\omega^{(#1)}}
\newcommand{\odp}{\braket{\braket{X;Y}}_\omega}
\newcommand{\odpk}{\braket{\braket{X;Y}}_{\omega_k}}
\newcommand{\rag}{\rangle}
\newcommand{\lag}{\langle}
\newcommand{\mt}[1]{\mbox{\tiny{#1}}}
\newcommand{\quadraom}{\braket{\braket{X;Y,Z}}_{\omega_Y, \omega_Z}}
\newcommand{\quadrak}{\braket{\braket{X;Y,Z}}_{\omega_{k_1}, \omega_{k_2}}}

\newcommand{\quadru}[3]{\frac{\braket{\Psi_0|#1|\Psi_K} [\braket{\Psi_K|#2|\Psi_N} -\delta_{KN}\braket{\Psi_0|#2|\Psi_0}]\braket{\Psi_N|#3|\Psi_0}}{(\omega_K +\omega_{#1})(\omega_N -\omega_{#3})}}

\newcommand{\etd}{e^{T^\dagger}}
\newcommand{\esd}{e^{S^\dagger}}
\newcommand{\esdm}{e^{-S^\dagger}}
\newcommand{\etdm}{e^{-T^\dagger}}
\newcommand{\esm}{e^{-S}}
\newcommand{\etm}{e^{-T}}
\newcommand{\es}{e^{S}}
\newcommand{\et}{e^{T}}

\newcommand{\omz}{\Omega^Z}
\newcommand{\omy}{\Omega^Y}

\newcommand{\Omx}{\Omega^X(\omega)}
\newcommand{\Omm}{\Omega(\omega)}

\newcommand{\karm}{\kappa(r_M, S, T)}
\newcommand{\kark}{\kappa(r_K, S, T)}
\newcommand{\karl}{\kappa(r_L, S, T)}
\newcommand{\etrn}{\eta(r_N, S)}
\newcommand{\etrl}{\eta(r_L, S)}
\newcommand{\etrm}{\eta(r_M, S)}

\newcommand{\xmz}{X_0}

\newcommand{\odpq}{\braket{\braket{X;Y,Z}}_{\omega_Y,\omega_Z}}

\newcommand{\phm}{\hat{\mathscr{P}}}
\newcommand{\pz}{\Psi_0}

\journal{Advances in Quantum Chemistry}

\begin{document}

\begin{frontmatter}

  \title{Molecular properties from the explicitly connected expressions of the response functions within the coupled-cluster theory}

  \author{Aleksandra M. Tucholska}
  \ead{am.tucholska@uw.edu.pl}
\author{Robert Moszynski}

\address{Faculty of Chemistry, University of Warsaw, Pasteura 
 1, 02-093 Warsaw, Poland}

\begin{abstract}
  We review  the methods based on   expectation value coupled cluster formalism - a common framework for the derivation of properties:
    the ground-state average value of an observable, cumulants of the
  second-order reduced density matrices, polarization propagator, quadratic
  response function, and  transition probabilities.
  We discuss the approximations and give examples of the most important numerical results.
\end{abstract}
\end{frontmatter}

\tableofcontents

\section{Introduction}
\subsection{Molecular properties}
The development of quantum chemical methods allows one to understand 
why physical phenomena occur and how to model them using computed physical properties of atoms and molecules: ionization energies, 
 electron affinities, multipole moments, polarizabilities, intermolecular forces, and transition properties.
 In the past decades numerous \textit{ab initio} quantum chemical methods were developed:
 the configuration interaction (CI)\cite{shavitt1998history}, M\o ller--Plesset perturbation theory (MP) also known
as the many-body perturbation theory (MBPT), coupled cluster theory (CC),\cite{vcivzek1966correlation, cizek1969use} 
Monte Carlo methods,\cite{motta2017computation, vrbik1990infinitesimal, thomas2015analytic}
and density-matrix renormalization group (DMRG)\cite{white1992density, white1993density, schollwock2005density}. Those approximations can be
employed as frameworks for the computation of the energy and properties
of many-electron systems.

In general, molecular properties can be classified as 
energy difference properties, response properties,
transition probabilities, internal interaction properties, internal structure properties and other kinds.\cite{pickup1992theory} 
Consider a system with a time-dependent perturbation $V(t)$. One can expand the 
time-dependent expectation value of an observable $X$ in orders of the perturbation $V(t)$. 
The resulting terms are the time-independent expectation value of $X$,  and linear, quadratic, cubic, and higher-order response functions.
Since the work of Oddershede,\cite{oddershede1978polarization, oddershede1984polarization, oddershede1987propagator}
who introduced the concept of response functions into molecular physics, 
 response theory has proven to be an  important tool for the calculation of 
molecular properties.\cite{linderberg2004propagators, jorgensen2012second, oddershede1987propagator}
This work focuses on  the
expectation value formalism of the coupled cluster theory (XCC) 
applied to  response properties and transition probabilities.

\subsection{Coupled cluster theory}
The coupled cluster theory\cite{coester1958bound, coester1960short, vcivzek1966correlation, cizek1969use} 
is the leading quantum-chemical wave function approach for 
 describing of the electronic structure of small and medium-sized systems.
The hierarchy of approximations in the CC theory provides an effective description
of the electron correlation while retaining the size consistency.\cite{dykstra2011theory}
The coupled cluster Ansatz is 
\equl{\Psi = e^T \Phi,}{ansatz}
where $\Phi$ is a reference determinant and $T$ is the coupled cluster excitation operator. 
Since the seminal works of Coester and Kümmel\cite{coester1958bound, coester1960short} and later \v{C}\'{\i}\v{z}ek, \cite{vcivzek1966correlation} numerous applications of the method were reported\cite{paldus1972correlation} and general-purpose programs\cite{pople1978electron, purvis1982full} were developed.  
CC is routinely used for the computation of correlated ground-state energies, \cite{bartlett2007coupled}
molecular properties,
\cite{koch1997coupled, korona2006time, korona2006one}
excited-state properties,\cite{krylov2008equation}
and analytical gradients.\cite{adamowicz1984analytical} It has been applied to atoms,
molecules and extended systems.\cite{hirata2001highly, hirata2004coupled, mihaila2000ground, kowalski2004coupled}

The cluster operator $T$ for an $N$ electron system is a sum of single, double, and higher excitations, $T=T_1+T_2+\cdots + T_{N}$. 
The $n$-particle excitation operator  $T_n$ is expressed as
\equ{
T_n = \frac{1}{n!^2}\sum_{\mu_n} t_{\mu_n} \mu_n,
}
where $t_{\mu_n}$ are the CC amplitudes, and $\mu_n$
\equ{
  \mu_n = e^{a_1a_2\ldots a_n}_{i_1i_2\ldots i_n}
}
denote the spin-orbital substitution operators $e^{p}_{q}$
\equ{
  \mu_1 = e^{a_1}_{i_1} = a_1^\dagger i_1 \qquad  \mu_2 = e^{a_1a_2}_{i_1i_2} = a_1^\dagger a_2^\dagger i_2 i_1 \qquad \ldots
}
$a_1^\dagger a_2^\dagger\ldots$ are  creation operators and $i_1i_2\ldots$ are annihilation operators.
The labels $a_1, a_2, \ldots i_1, i_2, \ldots p_1, p_2, \ldots$ are used
for the virtual, occupied and general spinorbital indices.
The ground-state energy is found by inserting \fr{ansatz} into the time-independent Schr\"{o}dinger equation, multiplying
 from left by $e^{-T}$, and projecting onto  the reference determinant. The amplitudes are obtained 
by projection onto excited determinants:
\equa{
&\lag \Phi|e^{-T}H e^{T}|\Phi\rag = E,\\
&\lag \mu_1|e^{-T}H e^{T}|\Phi\rag = 0,\nonumber\\
&\lag \mu_2|e^{-T}H e^{T}|\Phi\rag = 0,\nonumber\\
&\ldots\nonumber\\
&\lag \mu_N|e^{-T}H e^{T}|\Phi\rag = 0.\nonumber
}

For many-electron systems, one needs to introduce approximations to keep the method computationally feasible. The natural choice is to truncate
the $T$ operator at a given excitation level, obtaining  CCD\cite{vcivzek1966correlation}, CCSD\cite{purvis1982full, scuseria1988efficient, scuseria1987closed}, CCSDT,\cite{noga1987j, noga1988erratum, scuseria1988new}
etc. The next step is to neglect some terms in the amplitude
equations as in the  CC2\cite{christiansen1995second} and  CC3\cite{koch1997cc3} approximations.

In the  CCSD approximation, the  operator $T$ is truncated after the double excitations, i.e., $T = T_1 + T_2$.
The computational cost of the ground-state energy and amplitudes scales
as $N^6$, where $N$ is the size (number of electrons) of the system. 
In the  CC2\cite{christiansen1995second} method, the single amplitudes are computed in the same way as in CCSD.
 The double amplitudes are correct only through the first-order in the fluctuation potential  $W$
\equl{ W = H - F, 
}{www}
where $H$ is the Hamiltonian and $F$ is the Fock operator.
The CC2 approximation scales as $N^5$.

For the full inclusion of the triple amplitudes, the CCSDT model\cite{noga1987j, noga1988erratum, scuseria1988new} has to be used.
However, the computational cost,
 $N^8$, limits its applications.
Various approaches for approximate inclusion
of the triple excitations are available in the literature,
including CCSDT-1,\cite{lee1984coupled, urban1985towards, noga1987towards}
CCSD(T),\cite{raghavachari1989fifth} CC3\cite{koch1997cc3} and SVD-CCSDT.\cite{lesiuk2019implementation}
In the CCSD(T) model, it is unnecessary to  explicitly solve for $T_3$ and the triple excitations are  included only
in the energy expression as the fourth- and fifth-order perturbation energy contributions.
 However this method is not suitable for the computation for the transition properties because of the perturbative inclusion
 of triples. By contrast,
 in the CC3 method, $T_1$ and $T_2$ are
 iterated in the presence of approximate $T_3$. The computational cost of obtaining  $T_3$ in this method is $N^7$. The availability of iterative amplitudes
 makes CC3 a viable approach beyond-CCSD approach for the calculation of molecular properties. In the SVD-CCSDT method the scaling is reduced to $N^6$
 from the original $N^8$ scaling of CCSDT by the use of tensor decomposition. This method has been
 published only recently and is a promising approximation to be applied together with the XCC method.

\subsection{Response functions}
A system with a time-dependent perturbation $V(t)$ is described by the Hamiltonian
\equl{H(t) = H_0 + V(t),
}{ht}
where $H_0$ is the Hamiltonian of the unperturbed system and the perturbation can be written in a general, Hermitian
form as a sum of periodic contributions\cite{christiansen1998response} 
\equl{V(t) = \sum_{k=-M}^M e^{-i\omega_k t}\sum_y \epsilon_y (\omega_k) y
}{zab}
$\omega_k$ is the frequency of the perturbation and $\epsilon_y(\omega_k)$ is the perturbation strength parameter.
The following equalities hold for thus defined, Hermitian $V(t)$:
\equa{
  y^\dagger &= y\\
\omega_{-k} &= -\omega_k\\
\epsilon_y^*(\omega_k) &= \epsilon_y(-\omega_k).
}
The expectation value of an observable $X$ is
\equ{\bar{X}(t) = \braket{\Psi_0(t)|X|\Psi_0(t)}.}
 Formally $\bar{X}(t)$ is obtained by solving the time-dependent Schr\"{o}dinger equation 
\equl{H(t) \Psi_0(t) = i \frac{\partial}{\partial t} \Psi_0(t),
}{tds}
and expanding the expectation value  in orders of the perturbation
\equml{\bar{X}(t) = \bar{X} +\sum_{k} e^{-i\omega_k t}\sum_y \epsilon_y (\omega_k) \odpk +\\
  \frac12\sum_{k_1, k_2} e^{-i(\omega_{k_1}+\omega_{k_2}) t}\sum_{y,z} \epsilon_y (\omega_{k_1})\epsilon_z (\omega_{k_2}) \quadrak + \ldots
}{pert-exp}
The first term in this expansion, $\bar{X}$, is the time-independent expectation value. The coefficients $\odpk$, $\quadrak $, etc.
are the linear, quadratic and higher response functions, respectively, which 
describe the response of an observable $X$ to the perturbation $V(t)$.

There exist two approaches to the coupled cluster response theory. 
The first approach based on 
the differentiation of the CC energy expression was introduced by
Monkhorst in 1977\cite{monkhorst1977calculation, dalgaard1983some} and 
later extended by Bartlett et. al,\cite{jorgensen2012geometrical, fitzgerald1986analytic, salter1987property, salter1989analytic}
and is known as the $\Lambda$ operator technique.
Koch and J{\o}rgensen\cite{jo1988mo, helgaker1989configuration, koch1990coupled}
proposed the time-averaged quasi-energy Lagrangian technique, which is referred to in this work
as the time-dependent coupled cluster approach (TD-CC). That method requires the solution for the  Lagrange multipliers
in addition to coupled cluster amplitudes.
Within that approach, its authors developed expressions for the linear,
quadratic, and cubic response functions,\cite{christiansen1998response}
transition moments,\cite{christiansen1998integral} spin-orbit
coupling matrix elements,\cite{christiansen2002radiative, helmich2016spin}
and other properties,\cite{helgaker2012recent}
at the CC2,\cite{christiansen1995second, hald2000linear} CCSD,\cite{koch1994caclculation} and 
CC3 levels of theory.\cite{koch1997cc3, hald2002calculation}

The second approach is based on the computation of molecular properties
directly from  the average value of the operator $X$ using the auxiliary excitation operator $S$.
To avoid confusion it should be noted that the XCC method considered in this work is different from the approach of
Bartlett and Noga\cite{bartlett1988expectation} bearing the same name.
An extensive review of the $S$ operator method was published 
by Arponen and co-workers in the context of the extended coupled cluster theory (ECC).\cite{arponen1983variational, arponen1987extended, bishop1990correlations}
The  operator $S$ is defined in that work  by
a set of  nonlinear equations to which no systematic scheme of approximations had been found.
Later Jeziorski and Moszynski\cite{jeziorski1993explicitly} investigated this problem and
proposed an expression for $S$  that satisfies a set of  linear equations and can be systematically approximated. Within this approach the
average value of $X$ can be expressed through a finite series of commutators. As commutators cotain only contracted terms, the expression for
 the expectation value of $X$ is fully connected.
 This method was then extended by Moszynski et al. \cite{moszynski2005time} to the computation of the polarization propagator.

 To date, the XCC method\cite{jeziorski1993explicitly} has been employed to compute various electronic properties: electrostatic\cite{moszynski1993many} and exchange\cite{moszynski1994many}
contributions to the interaction energies of closed-shell systems, first-order molecular properties,\cite{korona2006one}  frequency-dependent density susceptibilities,\cite{korona2008dispersion} static and dynamic dipole polarizabilities, \cite{korona2006time}
and transition moments between
excited states.\cite{tucholska2014transition, tucholska2017transition} 
A prominent application of the XCC formalism
is the computation of symmetry-adapted perturbation theory (SAPT) contributions described in a series of publications by
Korona et al.\cite{korona2008first, korona2008second, korona2008dispersion, korona2009exchange} 
\section{Explicitly connected expansion of an observable's  average value}
The expectation value in the coupled cluster theory is defined as
\equl{ \bar{X} = \frac{\braket{\Psi|X\Psi}}{\braket{\Psi|\Psi}} = \frac{\braket{\etd X\et}}{\braket{\etd\et}},
}{r1}
where we use a shorthand notation which skips the Hartree-Fock determinant:
\equ{
  \braket{X} \equiv \braket{\Phi|X\Phi}
}
\equ{\ket{Xe^T} \equiv \ket{Xe^T\Phi}
  }
and
\equ{
  \braket{X|Y} \equiv \braket{X\Phi|Y\Phi}.
}
The expansion of \fr{r1} in $T$ is infinite 
due to 
the presence of the $\etd$ operator. However it can be reformulated with the use of an auxiliary operator $S$\cite{arponen1983variational, jeziorski1993explicitly}
\equl{e^S\Phi = \frac{\etd \et \Phi}{\braket{\etd\et}}.
}{r3}
as
\equl{ \bar{X} = \frac{\braket{\etd \et |\etm X \et}}{\braket{\etd\et}} = \braket{e^S|\etm X\et} = \braket{\esd\etm X \et\esdm}.
}{r2}
In the last equality we used identity b from \Frt{facts}.
\begin{table}[!bt]
  \begin{center}
    \caption{Identities used in the XCC derivations.}\label{facts}
  \begin{tabular}{l l}
    \toprule
a)&   $e^{-T}H e^{T} = H + \comute{H}{T} + \frac{1}{2!}\comute{\comute{H}{T}}{T}+ \ldots$\\[7pt]
   b)& $\ket{\esdm \Phi}=\Phi$\\[7pt]
    c)&$X\Phi = \braket{X}\Phi + \hat{\mathscr{P}}(X)\Phi$\\[7pt]
    d)&$1=\etm\esd\esdm\et$\\[7pt]
    \bottomrule
\end{tabular}
\end{center}
\end{table}
The auxiliary operator $S$  
\equl{S = S_1 + S_2 + S_3 +\ldots
}{r4}
is an excitation operator and satisfies a linear equation, where $S_n$ is expanded
 in a finite commutators series\cite{jeziorski1993explicitly}
\equal{
S_n &= T_n  - \frac1n \hat{\mathscr{P}}_n \left ( \sum_{k=1}
\frac{1}{k!}\cm{\widetilde{T}\dgg}{T}_k \right ) \\
&- \frac1n \hat{\mathscr{P}}_n\left (\sum_{k=1}\sum_{m=0}
\frac{1}{k!}\frac{1}{m!} [\cm{\widetilde{S}}{\dg{T}}_k,T]_m\right)\nonumber,
}{ss}
where
 \equ{
\widetilde{T} = \sum_{n = 1}^{N} nT_n \qquad\widetilde{S} = \sum_{n = 1}^{N} nS_n.
}
 $[A, B]_k$ is a shorthand for a $k$-times nested commutator.
 The superoperator $\hat{\mathscr{P}}_n(X)$ 
yields the 
excitation part of $X$
\equl{\hat{\mathscr{P}}_n(X) = \frac{1}{n!^2}\sum_{\mu_n} \braket{\mu_n|X}\mu_n.}{superp}
The  operator $S$ is explicitly connected.\cite{jeziorski1993explicitly}
Using the commutator expansion of $\etm X\et$ the expectation value takes the form
\equl{ \bar{X} = \braket{\esd\etm X \et \esdm} = \sum_{k=0}\sum_{m=0}\frac{1}{k!}\frac{(-1)^m}{m!}\braket{\left[\left[X, T\right]_k,S^\dagger\right]_m}.
}{r5}
Because $\bar{X}$  is expressed solely by the commutators of connected operators, it is 
explicitly connected. We assume 
that $X$ is a one-electron Hermitian operator. An arbitrary non-Hermitian operator can be
represented as a linear combination of $X+X^\dagger$ and $i(X-X^\dagger)$, which are both Hermitian.

Although the expression for $S_n$ is finite, it includes complicated terms with high powers of $T$, even when $T$ is truncated
and therefore needs to be approximated. The expectation value $\bar{X}$ from \fr{r5} is also approximated independently of $S$.
For example at the CCSD level the average value $\bar{X}$ becomes\cite{korona2006one}
\equsl{\bar{X} &= \braket{X} \\
  &+ \braket{S_1|X} + \braket{XT_1} + \braket{S_2|[X, T_2]}\\
  &+\braket{S_1|[X, T_2]} \\
  &+ \braket{S_1|[X, T_1]} +\braket{S_2|[[X, T_1], T_2]} \\
  &+\frac12\braket{S_1^2|[X, T_2]} +\frac12\braket{S_1S_2|[[X, T_2], T_2]} \\
  &+ \frac12\braket{S_1|[[X, T_1], T_1]} +\frac12\braket{S_3|[[X, T_2], T_2]} \\
  &+ \frac12\braket{S_1^2|[[X, T_1], T_2]} \\
  &+\frac{1}{12}\braket{S_1^3|[[X, T_2], T_2]}.
}{x-ccsd}
In general the approximations of \fr{r5} are guided by either MBPT expansion or expansion in powers of $T$.
\subsection{MBPT expansion}\label{sec-pert}
The $T$ amplitudes can be expanded in orders of $W$\cite{monkhorst1981recursive} 
\equl{ T= T^{(1)} + T^{(2)} + T^{(3)} + \ldots
}{t-orders}
The  amplitudes $T^{(n)}$, where $n$ is the MBPT order, are obtained with the  resolvent
\equl{\resnd{n}(W) = \frac{1}{(n!)^2}\sum_{a_ni_n}\frac{\braket{e^{i_1\ldots i_n}_{a_1\ldots a_n}W}e_{i_1\ldots i_n}^{a_1\ldots a_n}}
  {(\epsilon_{i_1}+ \ldots + \epsilon_{i_n} -\epsilon_{a_1}-\ldots -\epsilon_{a_n})} 
  }{res}
\equa{&T^{(1)} = T_2^{(1)} = \resnzd{2}(W)\\
  &T_n^{(2)} = \resnzd{n}([W,T_2^{(1)}]), \qquad n=1, 2, 3\\
  &T_1^{(3)} = \resnzd{1}([[W, T_1^{(2)} + T_2^{(2)} + T_3^{(2)}  ]).
  }
  The corresponding MBPT approximations to $S$ are
\equ{S^{(1)} = T^{(1)}}
\equ{S^{(2)} = T^{(2)}}
\equ{S_1^{(3)} = T_1^{(3)}+ \hat{\mathscr{P}}_1\left ([T_1^{(2)\dagger}, T_2^{\pr}] \right )  \\
+ \hat{\mathscr{P}}_1\left ([T_2^{(1)\dagger}, T_3^{\dr}] \right )}
\equ{S_2^{(3)} = T_2^{(3)} +\frac12\hat{\mathscr{P}}_2\left ([[T_2^{(1)\dagger}, T_2^{(1)}], T_2^{(1)}] \right).
  }
To obtain $\bar{X}$ at a given MBPT order, one takes only selected terms from
\fr{r5}
\equ{\bar{X}\dzr = \braket{X}}
\equ{\bar{X}\dpr = 0}
\equ{\bar{X}\ddr  =\braket{S_1\ddr|X} + \braket{XT_1\ddr} + \braket{S_2\dpr|[X,T_2\dpr]}}
\equm{\bar{X}\dtr = \braket{S_1\dtr|X} + \braket{XT_1\dtr} + \braket{(T_1\ddr+T_2\ddr+T_3\ddr )| [X, T_2\dpr]} + \\
 \braket{T_2\dpr|[X, (T_1\ddr+T_2\ddr+T_3\ddr )] }. 
}
\subsection{Expansion in  powers of $T$}\label{secnonp}
In practice, it is easier to apply an expansion in  powers of $T$.
The  $T$ amplitudes are first computed from the conventional CC theory. The
$S$ amplitudes in this approach are expanded in powers of $T$, giving the following
equations in the CCSD case
\equ{S^{[1]}_n = T_n}
\equ{S_1^{[2]} = \hat{\mathscr{P}}_1\left ([T_1^{\dagger}, T_2] \right )  }
\equ{S_2^{[3]} = \hat{\mathscr{P}}_2\left (\frac12[[T_2^{\dagger}, T_2],T_2]  + [[T_1^{\dagger}, T_2], T_1]\right )  }
\equ{S_1^{[3]} = \hat{\mathscr{P}}_1\left (\frac12[[T_1^{\dagger}, T_1],T_1]  + [[T_2^{\dagger}, T_2], T_1]\right )  }
\equl{S_3^{[3]} = \frac12\hat{\mathscr{P}}_3\left ([[T_1^{\dagger}, T_2],T_2]\right).}{drogi}
The superscript $[n]$ in the above equations denotes the $n$th power of $T$, and XCCSD[n] denotes the XCCSD method
comprising terms up to the  $n$th power of $T$. 
It is important to notice that even for the  CCSD case, there are $S_3$ amplitudes present.

Molecular properties calculated from \fr{r5} approximated at the CCSD level
 are only correct through  $\mathscr{O}(W^2)$ because the  terms
$\braket{S_1X}$, $\braket{XT_1}$  and  $\braket{T_2|[X, T_3]}$ 
introduce the $\mathscr{O}(W^3)$ error. 
Although \fr{x-ccsd} contains several terms that are of high
order in MBPT,   they do not introduce computational difficulties.\cite{korona2006one}  
The computational cost of the CCSD amplitudes scales as $\mathscr{O}(o^2v^4)$,
where $o$ stands for occupied and $v$ for virtual indices.
The $S$ amplitudes are computed afterwards in one step. The most expensive term is $S_3^{[3]}$ of \fr{drogi}.
Instead of computing $S_3^{[3]}$ first,
and $\braket{S_3^{[3]}|[[X, T_2], T_2]}$ in the separate step,
the authors of Ref~\citep{korona2006one}  proposed to
used  it directly as
\equ{\braket{S_3^{[3]}|[[X, T_2], T_2]} = \frac12\braket{[[T_1^\dagger, T_2], T_2]|[[X, T_2], T_2]}.
}
This allows for a factorization which reduces the 
 cost of evaluating the above term to $\mathscr{O}(o^3v^3)$.
The computational cost of the rest of the $S$ amplitudes and XCCSD[3]  scales as $\mathscr{O}(o^3v^3)$ or lower.
\begin{table}[!ht]
  \begin{adjustbox}{max width=\textwidth}
   \begin{threeparttable}
     \caption{Reprinted from Ref~\citep{korona2006one} with permission of AIP Publishing (Table II, page 184109-7).
       Contributions from the consecutive terms in \fr{x-ccsd} to the dipole
       moment of HF, CO, and BeO, and the quadrupole moment of HF, benzene, CO,
       and BeO. The calculations for HF, benzene, and CO were performed in the
       aug-cc-pVTZ basis, and for BeO in the Sadlej polarization basis. All values are
in a.u.}
     \label{tabb2}
\begin{tabular}{l c lllllll}
  &
  &   \multicolumn{1}{c}{HF}
    &   \multicolumn{1}{c}{HF}
    &   \multicolumn{1}{c}{Benzene}
    &   \multicolumn{1}{c}{CO}
    &   \multicolumn{1}{c}{CO}
    &   \multicolumn{1}{c}{BeO}
  &   \multicolumn{1}{c}{BeO}  \\
        \multicolumn{1}{l}{Terms of \fr{x-ccsd}}  
    &   \multicolumn{1}{c}{$\mathscr{O}(W^n)$}
    &   \multicolumn{1}{c}{$\mu$}
    &   \multicolumn{1}{c}{$\Theta$}
    &   \multicolumn{1}{c}{$\Theta$}
    &   \multicolumn{1}{c}{$\mu$}
    &   \multicolumn{1}{c}{$\Theta$}
    &   \multicolumn{1}{c}{$\mu$}
    &   \multicolumn{1}{c}{$\Theta$}\\
    \hline
1 & 0 & -0.7575 & $\,\,$1.7350 & -6.6446 & -0.1050 & -1.5460 & -2.9462 & $\,\,$5.3396\\
2,3,4 & 2 & $\,\,$0.0622 & -0.0300 & $\,\,$0.5961 & $\,\,$0.2277 & $\,\,$0.0735 & $\,\,$0.5971 & -0.1576\\
5 & 3 & -0.0029 & $\,\,$0.0025 & -0.0117 & -0.0169 & -0.0052 & -0.0430 & -0.0213\\
6,7 & 4 & $\,\,$0.0008 & $\,\,$0.0004 & -0.0003 & -0.0004 & -0.0065 & $\,\,$0.0916 & -0.1301\\
8,9 & 5 & -4$\cdot 10^{-6}$ & -2$\cdot 10^{-5}$ & -0.0002 & $\,\,$0.0001 & $\,\,$0.0010 & -0.0048 & $\,\,$0.0072\\
10,11 & 6 & -9$\cdot 10^{-7}$ & -4$\cdot 10^{-6}$ & -0.0002 & -8$\cdot 10^{-6}$ & $\,\,$0.0001 & -0.0012 & $\,\,$0.0008\\
12 & 7 & $\,\,$8$\cdot 10^{-7}$ & $\,\,$1$\cdot 10^{-8}$ & -1$\cdot 10^{-6}$ & $\,\,$3$\cdot 10^{-5}$ & -1$\cdot 10^{-6}$ & $\,\,$0.0003 & -2$\cdot 10^{-5}$\\
13 & 8 & -1$\cdot 10^{-9}$ & -1$\cdot 10^{-8}$ & -5$\cdot 10^{-8}$ & -5$\cdot 10^{-7}$ & -1$\cdot 10^{-7}$ & -1$\cdot 10^{-5}$ & -3$\cdot 10^{-6}$\\
Total XCCSD[3] &  & -0.6974 & $\,\,$1.7078 & -6.0251 & $\,\,$0.1056 & -1.4833 & -2.3062 & $\,\,$5.0387\\
$\bar{X}_{\mt{resp}}(3)$ &  & -0.7085 & $\,\,$1.7130 & -5.8638 & $\,\,$0.0653 & -1.4553 & -2.4616 & $\,\,$5.1133\\
LR-CCSD, nonrel. &  & -0.7062 & $\,\,$1.7109 & -5.9888 & $\,\,$0.0554 & -1.4784 & -2.4841 & $\,\,$5.1164\\
LR-CCSD, rel. &  & -0.7114 & $\,\,$1.7111 & -5.9828 & $\,\,$0.0273 & -1.4766 & -2.5604 & $\,\,$5.1464\\
LR-CCSD(T), nonrel. &  & -0.7010 & $\,\,$1.7009 & -5.8340 & $\,\,$0.0657 & -1.4831 & -2.3834 & $\,\,$4.9897\\
LR-CCSD(T), rel. &  & -0.7047 & $\,\,$1.7082 & -5.8552 & $\,\,$0.0497 & -1.4809 & -2.4369 & $\,\,$4.9980
\end{tabular}
   \end{threeparttable}
   \end{adjustbox}
 \end{table}

\Frt{tabb2} (Table II from Ref~\citep{korona2006one}) presents the importance of contributions from the consecutive terms in \fr{x-ccsd}. 
The authors of Ref~\citep{korona2006one} computed dipole and quadrupole moments of several molecules.
In the case of HF molecule, one can observe the reduction by an order of
magnitude of the consecutive contributions to the multipole moment with the increase of MBPT order.
For BeO the situation is more complicated. The analysis done
by Korona\cite{korona2006one} shows that the large contribution of $\mathscr{O}(W^4)$ comes mainly from term 6, which contains $S_1$ and $T_1$ amplitudes. This effect follows from the fact that the singles amplitudes are large in the BeO case and the CCSD method is not adequate for this molecule.
The numerical results for other molecular properties such as 
relativistic mass-velocity, one-electron Darwin terms, and electrostatic energies of van der Waals complexes can be found
in Ref~\citep{korona2006one}.

\subsection{CC3 approximation to the ground-state average value}
A formula for  $\bar{X}$ in the CC3 approximation
was  derived and investigated numerically by Tucholska et al.\cite{tucholska2014transition} 
Let the $T_1$, $T_2$ and $T_3$ be the CC3 amplitudes.
$S_n(m)$ denotes an approximation of $S_n$, where the MBPT contributions up to $m$th order
are fully included given the CC3 model of $T$. Therefore, we define
\equsl{
S_1(2) &= T_1   \\
S_1(3) &= S_1\dr +  \hat{\mathscr{P}}_1\left ([T_1^{\dagger}, T_2] \right )  
+ \hat{\mathscr{P}}_1\left ([T_2^{\dagger}, T_3] \right )   \\
S_1(4) &= S_1\tr +  
\hat{\mathscr{P}}_1\left ([[T_2^{\dagger}, T_1], T_2] \right )
+\frac12\hat{\mathscr{P}}_1\left ([[T_3^{\dagger}, T_2], T_2 ]\right )  \\
S_2(2) &= T_2  \\
S_2(3) &= S_2\dr + \frac12\hat{\mathscr{P}}_2\left ([[T_2^{\dagger}, T_2], T_2] \right )   \\
S_2(4) &=  S_2\tr  + \hat{\mathscr{P}}_2\left ([T_1^{\dagger}, T_3] \right )   \\
S_3(2) &= T_3   \\
S _3(4) &=  S_3\dr + \frac12\hat{\mathscr{P}}_3\left ([[T_1^{\dagger}, T_2], T_2] \right )  
+ \hat{\mathscr{P}}_3\left ([[T_2^{\dagger}, T_2], T_3] \right ).
}{przybl-s}
Using the definitions of $S_n$ of \fr{przybl-s}, we define the following approximations of the $S$ operator
\equsl{
&\mathrm{XCC3S(2)}: S = S_1(2) + S_2(2) + S_3(2)\\
&\mathrm{XCC3S(3)}: S = S_1(3) + S_2(3) + S_3(2)\\
&\mathrm{XCC3S(4)}: S = S_1(4) + S_2(4) + S_3(2).
}{xcc3s}
Now, for each of the approximations in \fr{xcc3s} one needs to select a subset of
commutator terms included in the formula for $\bar{X}$. This selection is
guided by the lowest MBPT order contributed by a given class of term. The corresponding
subsets of contributions are summarized in \Frt{Th-Fo-av}.
\begin{table}[!ht]
  \begin{center}
   \begin{threeparttable}
     \caption{Contributions to $\bar{X}$ in the CC3 approximation.}
     \label{Th-Fo-av}
\linespread{1.2}
\begin{tabular}{lc}
    \multicolumn{1}{c}{Contribution}
    &   \multicolumn{1}{c}{MBPT order}\\
  \hline
  $\braket{X}$ & 0\\
  $\braket{S_1|X}+ \braket{[X, T_1]} +\braket{S_2|[X, T_2]}$&2\\
  $\braket{S_1|[X, T_2]}+ \braket{S_2|[X, T_3]}$&3\\
   $\braket{S_1|[X, T_1]}+\braket{S_2|[[X, T_1], T_2]}+\braket{S_3|[X, T_3]}+\frac12\braket{S_3|[[X, T_2], T_2]}$&4\\
$\frac12\braket{S_1^2|[X, T_2]}+\frac12\braket{S_1S_2|[[X, T_2], T_2]}+\frac12\braket{S_1S_2|[X, T_3]}$&5\\
$\frac12\braket{S_1|[[X, T_1], T_1]}+\frac12\braket{S_1^2|[X, T_3]}$&6\\
$\frac12\braket{S_1^2|[[X, T_1], T_2]}$&7\\
$\frac{1}{12}\braket{S_1^3|[[X, T_2], T_2]}+\frac16\braket{S_1^3|[X, T_3]}$&8
\end{tabular}
   \end{threeparttable}
    \end{center}
\end{table}
The convergence of $\bar{X}$ with the MBPT terms
included was demonstrated numerically in Ref~\citep{tucholska2014transition} using
the example of the dipole moment of the HF molecule. \Frf{tab2} shows the  
 unsigned percentage error  in the dipole moment 
 relative to the experimental value 
 for each of the approximations.
\begin{figure}[!ht]
  \begin{center}
    \includegraphics[width=0.6\linewidth]{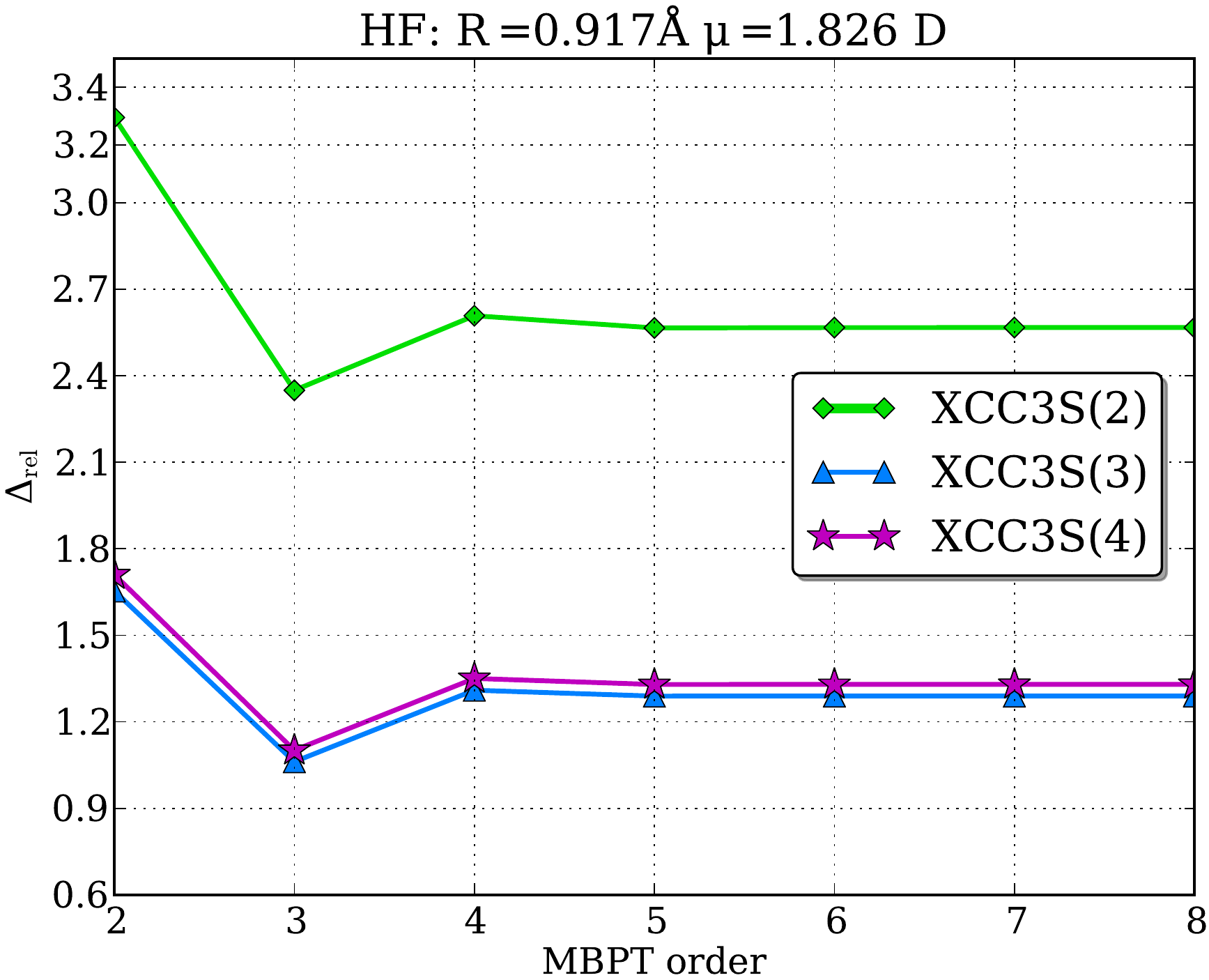}
    \caption{Reprinted from Ref~\citep{tucholska2014transition} with permission of AIP Publishing (Figure 1, page 124109-5).
      The dipole moment of the HF molecule. The  unsigned percentage error is defined as {\errname} $ = $ {\relerror}.
      MBPT order corresponds to \Frt{Th-Fo-av}.
      }
    \label{tab2}
    \end{center}
\end{figure}
For all three approximations of $S$, the convergence of $\bar{X}$
is achieved after inclusion of the 5th order terms.
To conclude, the convergence with respect to the $S$ operator approximation is achieved at the XCC3S(3) level and
within that approximation the commutator terms of \Frt{Th-Fo-av} need to be included up to the 5th order.
More numerical results for dipole and quadrupole moments and comparison to the experimental data
can be found in Ref~\citep{tucholska2014transition}.

\subsection{Ground-state two-particle density matrix cumulant}
The expression for the expectation value of an observable, \fr{r2}, can be further used to obtain the coupled cluster
parameterized density matrix cumulants.   
In the context of the coupled cluster theory, the cumulants appear for the first time in the work of Shork and Fuelde.\cite{schork1992derivation}
This subject was thoroughly studied some time later by Kutzelnigg and Mukherjee.
\cite{kutzelnigg1997normal, kutzelnigg1999cumulant}
Korona\cite{korona2008two} investigated the parametrization of the two-particle density
matrix cumulant employing the operator $S$.

The reduced one- and two-electron density matrix elements are defined in a given spinorbital basis as
\equa{&\Gamma(\bx_1| \bx_1') = \sum_{pq}\Gamma^p_q\phi_p^*(\bx_1)\phi_q(\bx_1')\\
  &\Gamma(\bx_1\bx_2| \bx_1'\bx_2') = \sum_{pqrs}\Gamma^{pr}_{qs}\phi_p^*(\bx_1)\phi_r^*(\bx_2)\phi_q(\bx_1')\phi_s(\bx_2')
}
where the expansion coefficients are
\equa{&\Gamma^p_q = \braket{\Psi|e^p_q \Psi}\\
  &\Gamma^{pr}_{qs} = \braket{\Psi|e^{pr}_{qs} \Psi}.
  }

The two-particle density matrix (2-RDM)  $\Gamma^{pr}_{qs}$ can be rewritten in the following way \cite{kutzelnigg1997normal}
\equl{\Gamma^{pr}_{qs} =     \Gamma^p_q\Gamma^r_s - \Gamma^p_s\Gamma^r_q  +\Lambda^{pr}_{qs} ,
  }{lambda}
where  the first two terms on the r.h.s. of the above equation are products of  one-electron density
matrices and the last term is the matrix element of the cumulant.
In contrast
to the 2-RDM, its cumulant is size-extensive.\cite{kutzelnigg1997normal}
Starting from the XCC formulation of the two-electron density matrix,\cite{jeziorski1993explicitly} we can write
\equ{\Gamma^{pr}_{qs} = \braket{\esd\etm e^{pr}_{qs}\et\esdm},
}
 using the relation\cite{paldus1988clifford}
\equ{e^{pr}_{qs} = e^p_qe^r_s -\delta_{rq}e^p_s,
}
and the equalities b and d of \Frt{facts}, 
we obtain the the expression
\equ{ \Gamma^{pr}_{qs} = \braket{\esd\etm e^p_q\et\esdm\esd\etm e^r_s\et\esdm} - \delta_{rq}\braket{\esd\etm e^p_s \et\esdm}.
}
This expression can by further reformulated 
to resemble \fr {lambda}
\equsl{\Gamma^{pr}_{qs} &= \braket{\esd\etm e^p_q \et\esdm} \braket{\esd\etm e^r_s \et\esdm}\\ &-
  \braket{\esd\etm e^p_r \et\esdm}\braket{\esd\etm e^q_s\et\esdm} \\
  &+ \braket{\esd\etm e^p_q \et\esdm \hat{\mathscr{P}}(\esd\etm e^r_s \et\esdm)}\\
  &-\braket{\esd\etm \tilde{e}^r_q\et\esdm}\braket{\esd\etm e^p_s\et\esdm},
}{gam1}
where  $\tilde{e}^r_q$  are one-hole substitution operators.\cite{paldus1988clifford} 
One can recognize the first two terms of \fr{gam1} as the products of one-electron density matrix elements present in \fr{lambda}.
The last two terms of \fr{gam1}, define the XCC cumulant of the two-electron density matrix
\equml{\Lambda^{pr}_{qs} = \braket{\esd\etm e^p_q \et\esdm \hat{\mathscr{P}}(\esd\etm e^r_s \et\esdm)}\\
  -\braket{\esd\etm \tilde{e}^r_q\et\esdm}\braket{\esd\etm e^p_s\et\esdm}.
}{lost}

In contrast to the expression for the XCC expectation value,
\fr{r5}, it is not obvious that  the XCC cumulant $\Lambda^{pr}_{qs}$ is connected.
Korona proved\cite{korona2008two} that some parts of the first term in \fr{lost}
cancel out the disconnected term. Therefore, we can write
\equ{\Lambda^{pr}_{qs} = \braket{\{\esd\etm e^p_q \et\esdm \hat{\mathscr{P}}(\esd\etm e^r_s \et\esdm)\}_C},
}
where the subscript $C$ means that one has to consider only the connected terms. 

The XCC approach was employed in reference SAPT theory\cite{jeziorski1994perturbation, szalewicz2005intermolecular}
calculations of Korona et. al.\cite{korona2008first}
A modified version of the first-order exchange energy formula\cite{moszynski1994many} of SAPT was
divided into two parts. The first one was expressed through
the one-electron density matrices and the second part through the monomer cumulants.
In this method, both monomers are described  at the CCSD level. 
Numerical results and extensive discussion can be found in the paper by Korona.\cite{korona2008first}

\section{Time-independent coupled cluster theory of the polarization propagator}\label{tpp}
In this section we  concentrate on the linear coefficient $\odp$ in the expansion of the time-dependent 
 expectation value of the operator $X$ from \fr{pert-exp}.
The linear response function is also known as the 
polarization propagator and was derived in  time-independent approach with the use of the auxiliary operator $S$
by Moszynski et. al.\cite{moszynski2005time}
The polarization propagator can be written down as\cite{oddershede1978polarization}
\equl{\odp = -\braket{\Psi_0|Y\frac{Q}{H - E_0 +\omega}X\Psi_0}
-\braket{\Psi_0|X\frac{Q}{H - E_0-\omega}Y \Psi_0},
}{1ord}
where $Q=1-\ket{\Psi_0}\bra{\Psi_0}$ is the projection operator on the space spanned by all excited states.
For Hermitian operators X and Y the second term in the above equation can be obtained by computing the first term for $-\omega^*$. This operation,
called the \textit{generalized complex conjugation}, 
is usually denoted as  \textit{g.c.c.} The propagator exhibits the following Hermitian symmetry
\equl{\braket{\braket{X;Y}}_\omega^*=\braket{\braket{X;Y}}_{-\omega^*}.}{tr}
Introducing the first-order wavefunction $\Psi^{(1)}(\omega)$
\equ{(H - E_0-\omega)\Psi^{(1)}(\omega)=-X\Psi_0
  }
the polarization propagator can be rewritten as
\equl{
\langle\langle X;Y\rangle\rangle _\omega = -\braket{\Psi_0|Y-\bar{Y}|\Psi^{(1)}(\omega)}+ \text{g.c.c.}
}{1ord}
$\Psi^{(1)}(\omega)$  parametrized with the excitation operator   $\Omega(\omega)= \Omega_1(\omega) + \Omega_2(\omega) + \ldots$, takes the form
\equl{|\Psi^{(1)}(\omega)\rangle =  (\Omega_0(\omega) + \Omega(\omega)) |\Psi_0\rangle.
}{psi1}
$\Omega_0(\omega)$ is a scalar term that ensures the orthogonality of $\Psi^{(1)}$ to $\Psi_0$\cite{moszynski2005time, korona2006time}
\equ{\Omega_0(\omega) = - \bb{\Psi_0}{\Omega(\omega)\Psi_0},}
and the amplitudes $\Omega_n(\omega)$   are the solutions of  the linear-response equation
\cite{moszynski2005time, korona2006time} 
\equl{
\left \langle \mu_n |\cm{e^{-T} H e^T}{ \Omega(\omega)} + \omega\Omega(\omega)
  + e^{-T}Xe^T\right \rangle = 0.
}{om}

Upon inserting the CC parametrized $\Psi^{(1)}(\omega)$  and
normalized CC ground state wavefunction $\Psi_0$\cite{moszynski2005time} 
  \equl{\Psi_0 = \frac{\et}{\braket{\etd\et}^{\frac12}}\Phi, }{e2}
the first term on the r.h.s. of \fr{1ord} becomes
\equl{\braket{\Psi_0|(Y-\bar{Y})\Psi^{(1)}(\omega)} =
  \frac{\braket{\et|Y|\Omega(\omega)\et}}{\braket{\et|\et}}
- \bar{Y}\frac{\braket{\et|\Omega(\omega)\et}}{\braket{\et|\et}},
  }{p2}
and using the formulas b and d from \Frt{facts} one arrives at
\equml{\braket{\Psi_0|(Y-\bar{Y})\Psi^{(1)}(\omega)} = \\
  \braket{\esm\etd Y \etdm\es|\esd\Omm)\esdm}\\
  - \braket{\esm\etd Y\etdm\es}\braket{\esd\Omm\esdm}.}{p3}
Similarly as in the case of the cumulant, the last term in the above formula  is explicitly 
disconnected and therefore some parts of the first term  cancel
out the second term on the r.h.s. This is demonstrated by using formula c from \Frt{facts} in the work of Moszynski et. al.\cite{moszynski2005time}
so the polarization propagator takes the form
\equl{
  \langle\langle X;Y \rangle \rangle_\omega =
  \braket{e^{-S}e^{T^\dagger}Ye^{-T^\dagger}e^{S}|\hat{\mathscr{P}}\left(e^{S^\dagger}\Omm e^{-S^\dagger}\right)} + \text{g.c.c.}
}{lin}
The XCC formulation of the linear response function is size-consistent as it is expressed solely in terms
of commutators. It is worth noting that the XCC polarization propagator exhibits the Hermitian symmetry, \fr{tr},
in contrast to the TD-CC polarization propagator. The \textit{a posteriori}
symmetrization\cite{christiansen1998integral, pedersen1997coupled}  of the latter recovers the missing term with
the wrong coefficient.\cite{moszynski2005time, korona2006time} 
\subsection{MBPT expansion}
Moszynski and Jeziorski\cite{moszynski2005time} proposed both perturbative and
non-perturbative schemes of approximating  $\odp$ 
within the CCSD model.
  In the former approach  the 
  $T$ and $S$ amplitudes are found in the same way as in \frs{sec-pert}. The order-by-order
  expressions for $\Omega_n$ are computed from the linear response equation 
  which can be rewritten as
\equl{\Omega_n(\omega) = \hat{\mathscr{R}}_{n,\omega}(\etm X\et + [\etm W\et, \Omega(\omega)]),
}{om-rzedy}
where
\equl{\resn{n}(W) = \frac{1}{(n!)^2}\sum_{a_ni_n}\frac{\braket{e^{i_1\ldots i_n}_{a_1\ldots a_n}W}e_{i_1\ldots i_n}^{a_1\ldots a_n}}
  {(\epsilon_{i_1}+  \ldots + \epsilon_{i_n} -\epsilon_{a_1}-\ldots -\epsilon_{a_n})-\omega}.
}{res2}
is the frequency-dependent resolvent.
Therefore, in the first two orders of MBPT, the contributions to  $\Omega_n$ are given by 
\equal{\Omega_1^{(0)} = &\resn{1}(X)\\
  \Omega_n^{(1)} = &\resn{n}([X,T_2^{(1)}] + [W, \Omega_1^{(0)}(\omega)]), \qquad n=1, 2\nonumber\\
  \Omega_n^{(2)} = &\resn{n}([X,T_1^{(2)} + T_2^{(2)}] + \nonumber\\
  &[W, \Omega_1^{(1)}(\omega)+ \Omega_2^{(1)}(\omega)]
  +[[W, T_2^{(1)}], \Omega_1^{(0)}(\omega)]
  ), \qquad n=1, 2, 3\nonumber
}{pp1}
and the polarization propagator itself is given by
\equal{&\odpr{0} = \braket{Y|\Omega^{(0)}(\omega)}+ \mbox{g.c.c.}\\
  &\odpr{1} = \braket{Y|\op{\Omega}{1}{1}(\omega)} + \braket{[Y,\op{T}{2}{1}]|\op{\Omega}{1}{0}(\omega)}+ \mbox{g.c.c.}\nonumber\\
  &\odpr{2} = \braket{Y|\op{\Omega}{1}{2}(\omega)} + \braket{[Y,\op{T}{2}{1}]|\op{\Omega}{1}{1}(\omega)}\nonumber\\
  & + \braket{[Y,\op{T}{2}{1}]|\op{\Omega}{2}{1}(\omega)}+ \braket{[Y,\op{T}{1}{2}]|\op{\Omega}{1}{0}(\omega)}\nonumber\\
  &+ \braket{[Y,\op{T}{2}{2}]|\op{\Omega}{1}{0}(\omega)} - \braket{[[Y,T_2^{(1)\dagger}],\op{T}{2}{1}]|\op{\Omega}{1}{0}(\omega)} + \mbox{g.c.c.}\nonumber
}{pp2}
\subsection{Expansion in powers of $T$}
In the non-perturbative approximation,
the $T$ amplitudes are computed from the conventional CC method, and the $S$ amplitudes are expanded in powers of $T$
in the same way as in \frs{secnonp}. The solution of \fr{om} for the $\Omega(\omega)$ operator is a standard procedure used to compute coupled
cluster gradients.\cite{salter1989analytic, korona2002electrostatic, scheiner1987analytic} Finally, the expression for the $\odp$ itself is given through the combined powers of $T$ and $S$ operators. In the CCSD case one obtains 
\equa{&\odps{0} = \braket{Y|\Omega_1(\omega)}+ \mbox{g.c.c.}\\
  &\odps{1} = \braket{S|[Y, \Omega(\omega)]} + \braket{[Y,S_2]|\Omega_1(\omega)}+ \mbox{g.c.c.}\nonumber\\
  &\odps{2} = -\braket{[[Y, T^\dagger], S]|\Omega(\omega)} + \braket{[Y, S]|[S_1^\dagger, \Omega_2(\omega)]} \nonumber\\
  &+ \frac12\braket{[[Y, S], S]|\Omega(\omega)} + \mbox{g.c.c.}\nonumber\\ 
  &\odps{3} = -\frac12\braket{[[[Y, T^\dagger], S], S]|\Omega(\omega)} - \braket{[[Y, T^\dagger], S]|[S_1^\dagger, \Omega_2(\omega)}\nonumber\\
      &+ \frac12\braket{[[[Y, T_1^\dagger],T_1^\dagger], S_2 |\Omega_1(\omega)} + \frac12\braket{[[Y, S_1], S_1]|[S_1^\dagger, \Omega_2(\omega)} + \mbox{g.c.c.}\nonumber
  }
The XCC method for the polarization propagator  was 
implemented by Korona et al.\cite{korona2006time, korona2006one, korona2008dispersion}
and a highly-optimized implementation is available
in the Molpro software package.\cite{molpro2015} In the XCC algorithm
 $\Omega(\omega)$ is obtained in an iterative procedure, and each iteration scales as $\mathscr{O}(o^2v^4)$. The calculation of
$\Omega(\omega)$ is the most expensive part. Computation of $S$ amplitudes and the polarization propagator, are one-step procedures, and
scale up to $\mathscr{O}(o^3v^3)$. In TD-CC method one must also compute the $\Lambda$ state, which is of similar cost as computing $\Omega(\omega)$.
Therefore this method is twice as expensive than XCC method.

As the first application the polarization propagator was used to compute static and
dynamic electric dipole polarizabilities\cite{korona2006time} of various molecules such as H$_2$, He, Be, BH, CH$^{+}$, Ne and HF.
The results were compared with the FCI method and it was confirmed numerically,
that the method is almost exact for
two-electron systems.\cite{moszynski2005time, korona2006time} The only approximation
originates from an inexact  treatment of the $S$ amplitudes, where
some of the multiply-nested commutators were neglected. 

The XCC formulation of the polarization propagator was also used to compute
the second order dispersion energy $E^{(2)}_{\mt{disp}}$ within the SAPT framework.\cite{korona2008dispersion}
The  Longuet-Higgins  formula for 
\equ{E^{(2)}_{\mt{disp}}  = -\frac{1}{2\pi}\int_0^\infty\int\int\int\alpha^A(\mathbf{r}_1, \mathbf{r}_2|i\omega)
  \alpha^B(\mathbf{r}_3, \mathbf{r}_4|i\omega) \frac{1}{r_{13}}\frac{1}{r_{24}}d\mathbf{r}_1d\mathbf{r}_2d\mathbf{r}_3d\mathbf{r}_4
  }
is expressed through frequency-dependent density susceptibilities
\equl{
  \alpha^A(\mathbf{r}_1, \mathbf{r}_2|\omega) = \braket{\braket{\hat{\rho}(\mathbf{r}_1);\hat{\rho}(\mathbf{r}_1)}}_\omega.
}{susc}
One can show that $\alpha^A(\mathbf{r}_1, \mathbf{r}_2|i\omega)$  is
obtained from  the polarization propagator $\odp$ where both $X$ and $Y$ are the electron density
operators
\equl{\hat{\rho}(\mathbf{r}_1) = \sum_{i=1}^N\delta(\mathbf{r}-\mathbf{r}_i).
}{dens}
In order to lower the computational cost, Korona and Jeziorski\cite{korona2008dispersion} used
the density fitting method\cite{whitten1973coulombic, dunlap1979some, kendall1997impact} for the computation
of the density susceptibilities. 

\subsection{First-order density matrix cumulant}
The XCC formalism was adapted for the computation of the second-order  exchange-induction energy of SAPT theory.\cite{korona2008second}
The $E_{\mt{exch-ind}}^{(2)}$  is defined in the $S^2$ approximation, as\cite{chalasinski1977exchange}
\equl{E_{\mt{exch-ind}}^{(2)} = \braket{\Psi_A\Psi_B|(V-\bar{V})(P-\bar{P})(\Psi^{(1)}_A\Psi_B + \Psi_A\Psi^{(1)}_B)},
  }{exch}
where $V$ is the intermolecular potential, and $P=-\sum_{i=1}^{N_A}\sum_{j=1}^{N_B}P_{ij}$ is a single-exchange operator.
  The authors of Ref~\citep{korona2008second} express $E_{\mt{exch-ind}}^{(2)}$  through 1-RDM and 2-RDM of monomers and subsequently divide the resulting expression in 
   the antisymmetrized product of 1-RDMs part and the cumulant part.
  
The first-order density matrix element is defined as
\equl{(\Gamma^{(1)})^{pr}_{qs} = \braket{\Psi_0|e^{pr}_{qs}\Psi^{(1)}}.
  }{e1}
To obtain the coupled cluster parametrization of this quantity, one uses the $\Psi_0$ and $\Psi_1$, \fr{e2} and \fr{psi1}, and takes $\omega=0$.
Inserting  the definition of the $S$ operator \fr{r3},
gets\cite{korona2008second}
\equl{\braket{\Psi_0|X\Psi^{(1)}} = \Omega_0 \braket{\esd\etm  X \et} + \braket{\esd\etm X \Omega\et}.
}{p1}
Further steps require to use the formulas from \Frt{facts} 
\equml{\braket{\Psi_0|X\Psi^{(1)}} = -\braket{\esd\Omega\esdm} \braket{\esd\etm  X \et\esdm} + \\
  \braket{\esd\etm X \et\esdm}\braket{\esd\Omega\esdm}
  +\braket{\esd\etm X \et\esdm\hat{\mathscr{P}}(\esd\Omega\esdm)},
}{p2}
where in first term  $\Omega_0$ is replaced with $-\braket{\esd\Omega\esdm}$.
The first-order density matrix becomes
\equl{(\Gamma^{(1)})^{pr}_{qs} = \braket{\Psi|e^{pr}_{qs}\Psi^{(1)}} = \braket{\esd\etm e^{pr}_{qs} \et\esdm\hat{\mathscr{P}}(\esd\Omega\esdm)}.
}{pp3}
The first-order cumulant $\Lambda^{(1)}$ of the first-order 2-RDM is defined as
\equml{(\Gamma^{(1)})^{pr}_{qs} =     
  (\Gamma^{(1)})^p_q(\Gamma^{(0)})^r_s +(\Gamma^{(0)})^p_q(\Gamma^{(1)})^r_s  \\-(\Gamma^{(1)})^p_s(\Gamma^{(0)})^r_q
  - (\Gamma^{(0)})^p_s(\Gamma^{(1)})^r_q
  +(\Lambda^{(1)})^{pr}_{qs} .
}{lambdafirst}
Korona\cite{korona2008second} showed how to divide \fr{pp3} in order to identify the first-order cumulant.
The formula c from \Frt{facts} is inserted between any pair of  operators so  the \fr{p3} becomes
\equm{(\Gamma^{(1)})^{pr}_{qs} = \langle\esd\etm p^\dagger\et\esdm\esd\etm r^\dagger \et\esdm\esd\etm q \times \\
  \times\et\esdm\esd\etm s \et\esdm \proj(\esd\Omega\esdm)\rangle,
}
and introducing the shorthand notation
\equa{
  &\Theta = \proj(\esd\Omega\esdm)\\
  &K^p=\esd\etm p^\dagger \et\esdm\nonumber \\
  &R_q=\esd\etm q \et\esdm\nonumber
}
one obtains an expression for $(\Gamma^{(1)})^{pr}_{qs}$ 
\equml{(\Gamma^{(1)})^{pr}_{qs} = \braket{\{K^pK^rR_sR_q\Theta\}_C}
  +\braket{\kp\rqq}\braket{\kr\rs\Theta}+\\
  \braket{\kr\rs}\braket{\kp\rqq\Theta} - \braket{\kr\rqq}\braket{\kp\rs\Theta}
   -\braket{\kp\rs}\braket{\kr\rqq\Theta},
}{gam1b}
where subscript C denotes taking only the connected part of the term.
The cumulant is identified as the first term on the r.h.s. 
of the above equation
\equ{(\Lambda^{(1)})^{pr}_{qs} = \braket{\{K^pK^rR_sR_q\Theta\}_C},
}
by comparison of the \fr{gam1b} and \fr{lambdafirst}.
This approach allows one to include the intramonomer correlation at the CCSD level for the many-electron monomers. 

\subsection{Transition moments from ground to excited states}
The natural step to further develop the XCC theory is to derive the expressions for the transition moments.
Transition moments from ground to excited electronic states can be found as a residue of the polarization propagator. 
In the exact theory, one can write the transition moments between the ground and an excited state $K$ as
\equ{
\lim_{\omega \rar \omega_K}(\omega - \omega_K)\odp
= \braket{\Psi_0|X \Psi_{K}}\braket{\Psi_{K}|Y\Psi_0} = T_X^{0K}T_Y^{K0}.
}
The product $T_X^{0K}T_Y^{K0}$ is called the transition strength matrix and satisfies the condition
\equl{\SSS_{XY}^{0K} = -(\SSS_{XY}^{K0})^*,
}{time-rev}
which is a direct consequence of the Hermitian symmetry of the polarization propagator.\cite{moszynski2005time}

In the XCC theory the excited states are obtained from the equation of motion coupled cluster theory (EOM-CC)
using CC Jacobian,\cite{sekino1984linear, koch1990coupled, helgaker2013molecular}
which  has the following form
\equl{A_{\mu_n\nu_m} =\frac{d \braket{\mu_n|e^{-T}H e^T|\Phi}}{d\nu_m} =  \braket{\mu_n|e^{-T}[H,\nu_m]e^T|\Phi}.
}{jac}
Due to the non-symmetric character of the Jacobian matrix, the diagonalization of $A$ leads to a
set of biorthogonal left $l_M$ and right $r_K$ eigenvectors
\equ{\braket{l_M|r_K} = \delta_{MK}.
  }
The transformation to the Jacobian basis is given by
\equ{\mu_n = \sum_{N}L^\star_{\mu_n \mbox{\tiny{\emph{N}}}} r_N,
  }
and is used to express the operator $\Omega(\omega)$ in terms of the Jacobian eigenvectors
\equl{
\Omx = \sum_M \sum_n\sum_{\nu_n} L_{\nu_n M}^*O_{\nu_n}^X(\omega) r_{M} = \sum_{M} O_{M}^X(\omega) r_{M}.
}{om-jak}
Inserting \fr{om-jak} to the linear response equation for $\Omm$, \fr{om}, one obtains
\equl{
\sum_{M}\braket{l_N|e^{-T}\comute{ H }{r_M}e^T}O_M^X(\omega) + \sum_{M}\omega\braket{l_N|r_M}O_M^X(\omega)
  + \braket{l_N|e^{-T}Xe^T} = 0,
}{raz}
where we use the superscript $X$ to indicate the $X$ dependence.
\fr{raz} can be simplified to the following form using the facts that the Jacobian matrix is diagonal in this basis,
and that this basis is biorthonormal
\equl{
\braket{l_N|e^{-T}\comute{ H }{r_N}e^T}O_N^X(\omega) +
\omega O_N^X(\omega)
  + \braket{l_N|e^{-T}Xe^T} = 0.
}{jakobian}
Denoting $\braket{l_N|e^{-T}\comute{ H }{r_N}e^T}$ as the nth excitation energy  $\omega_N =E_N - E_0 $, \fr{jakobian} can be rewritten as
\equ{
\omega_NO_N^X(\omega) +
\omega O_N^X(\omega)
  + \xi_N^X= 0
}
where 
\equl{\xi_N^X = \braket{l_N|e^{-T}Xe^T}.
}{xin}
The quantity $O_N^X(\omega)$ is then straightforwardly obtained as
\equl{
O_N^X(\omega) = - \frac{\xi_N^X }{\omega_N + \omega}.
}{om-jac}
\fr{lin} is  transformed to the Jacobian basis according to the following steps
\equml{
\langle\langle X;Y\rangle\rangle_\omega =
\braket{e^{-S}e^{T^\dagger}Ye^{-T^\dagger}e^{S}|\proj(e^{S^\dagger}\Omx e^{-S^\dagger})} + \mbox{g.c.c.} = \\
\sum_{n=1}\sum_{\mu_n}O^X_{\nu_n} \braket{e^{-S}e^{T^\dagger}Ye^{-T^\dagger}e^{S}|\proj(e^{S^\dagger}\tau_{\mu_n} e^{-S^\dagger})} + \mbox{g.c.c.} = \\
\sum_{N}O^X_{N} \braket{e^{-S}e^{T^\dagger}Ye^{-T^\dagger}e^{S}|\proj(e^{S^\dagger}r_N e^{-S^\dagger})} + \mbox{g.c.c.} 
}{trz}
Using \fr{om-jac} and
denoting
\equl{\gamma_{N}^Y = \braket{e^{-S}e^{T^\dagger}Ye^{-T^\dagger}e^{S}|\proj(e^{S^\dagger}r_N e^{-S^\dagger})},
  }{gamma3}
the XCC linear response function in the Jacobian eigenvector basis  becomes
\equl{
\langle\langle X;Y\rangle\rangle_\omega =
-\sum_N \frac{\xi_N^X \gamma_N^Y }{\omega_N + \omega}+ \mbox{g.c.c.}
}{odp-jak}
The poles of
\fr{odp-jak} are located at the
EOM-CC\cite{stanton1993equation} excitation energies, and 
the residue of the XCC response function reads 
\equ{
\lim_{\omega
  \rar -\omega_K}(\omega + \omega_K)\odp=
 -\underbrace{\gamma_{K}^Y}_{ T_Y^{0K}}
\underbrace{\xi_{K}^X}_{T_X^{K0}} =  \SSS_{XY}^{0K}.
}

The line strength  $\SSS_{XY}^{0K}$
is exclusively expressed in terms of commutators, 
and is automatically size consistent, regardless
of any truncation of the $T$ and $S$ operators.
The XCC transition strength satisfies the antihermiticity relation \fr{time-rev}
for any truncation of the operator $T$. The situation is more complicated for the truncated operator $S$.
In the derivations  we have used the formal definition of the operator
$S$,  $e^S\Phi = \frac{\etd\et}{\braket{\et|\et}}\Phi$, which is true only for the exact $S$.
Nonetheless,  the deviations from the exact Hermitian
symmetry are numerically negligible.\cite{tucholska2014transition, tucholska2017transition} This issue is discussed in more detail in the next section. 
The numerical results computed by Tucholska et al.\cite{tucholska2014transition} covers the dipole and
quadrupole transition probabilities for the Mg, Ca, Sr and Ba atoms, comparison to the TD-CC method and experimental data.
 \section{Quadratic response function}
 The next coefficient in the  expansion given by \fr{pert-exp} is the 
quadratic response function, $\braket{\braket{X;Y,Z}}_{\omega_Y, \omega_Z}$. It describes
a response of an observable $X$
to the perturbations $Y$ and $Z$ acting with the frequencies $\omega_Y$ and $\omega_Z$, respectively, and is defined as 
 the following sum over states \cite{christiansen1998response}
\equal{\langle\langle X; Y, Z \rangle \rangle&_{\omega_Y, \omega_Z}=\\[10pt]
=P_{XYZ}&\sum_{\substack{K=1\\N=1}}\quadru{Y}{X}{Z}\nonumber,
}{quadrubig}
where the operator $P_{XYZ}$ denotes the sum of all  permutations of the indices $X$, $Y$ and $Z$. 
$K$ and $N$ run over all possible excited states with the excitation energies $\omega_K$ and $\omega_N$ and $\Psi_0$ is the ground state.
By using the first-order perturbed wave function, \fr{psi1}, we rewrite \fr{quadrubig} as
\equl{ \odpq = P_{XYZ}\langle \Psi^{(1)}(\omega_Y)|\xmz|\Psi^{(1)}(-\omega_Z) \rangle,}{podst}
where we introduced the symbol $X_0 = X-\braket{X}$. In the form parametrized by CC wavefunctions, the quadratic
response function  reads
\equml{\quadraom   = \\P_{XYZ} \langle (\Omega_0(\omega_Y) + \Omega(\omega_Y))\pz|\xmz|(\Omega_0(\omega_Z) + \Omega(-\omega_Z))\pz\rangle.
}{start}
After expanding  \fr{start} and  using the 
formulas from \Frt{facts}, we get
 the final expression for the quadratic response function in XCC formalism
\equml{\odpq = \\P_{XYZ}\left(\bb{\phm(\esd \omy\esdm) }{ \esm\etd X_0 \etdm\es \phm(\esm\etd\omz \etdm\es)}\right).}{quadru1}

\subsection{Transition moments between  excited states}
A transition moment between two excited states is obtained from the residue of the quadratic response function. For
the exact wavefunction the residue is
\equal{&\lim_{\omega_Y\rightarrow -\omega_L}(\omega_L + \omega_Y)\lim_{\omega_Z\rightarrow \omega_M}(\omega_M - \omega_Z) \odpq\\
 = &\braket{\Psi_0|Y|\Psi_L}\braket{\Psi_L|X_0|\Psi_M}\braket{\Psi_M|Z|\Psi_0} =\TT^Y_{0L}\TT^X_{LM}\TT^Z_{M0}\nonumber.
}{tmexact}
By expanding \fr{quadru1} in the basis of the CC Jacobian eigenvectors \fr{om-jak}, one gets
\equa{&\quadraom = 
  P_{XYZ}\sum_{K, N =1}\left(O_K^Y(\omega_Y)\right)^{\star}O_N^Z(-\omega_Z)\\
&\times\bb{\phm(\esd r_K\esdm) }{ \esm\etd X_0 \etdm\es \phm(\esm\etd r_N \etdm\es)}.\nonumber
}
\normalsize
By introducing a shorthand notation for the projected parts of the above expression
\equs{
&\karm = \phm\left(\esm\etd r_K\etdm\es\right)\\
&\etrn = \phm\left(\esd r_N\esdm\right),
}
one arrives at the final form of the XCC quadratic response function in the Jacobian eigenvector basis
\equa{&\quadraom = \\
  &P_{XYZ}\sum_{\substack{K=1\\N=1}}\left(O_K^Y(\omega_Y)\right)^{\star}O_N^Z(-\omega_Z)
  \braket{\kark|\esd\etm \xmz\et\esdm \etrn }\nonumber\\
&= P_{XYZ}\sum_{\substack{K=1\\N=1}}\frac{\braket{\etm Y\et|l_K}}{\omega_K + \omega_Y}\frac{\braket{l_N|\etm Z\et}}{\omega_Z - \omega_N}
  \braket{\kark|\esd\etm \xmz\et\esdm | \etrn }.\nonumber}
The  residue of the  XCC quadratic response function reads
\equal{\TT_{0L}^Y &\TT_{LM} \TT^Z_{M0} = \lim_{\omega_Y\rightarrow -\omega_L}(\omega_L + \omega_Y)\lim_{\omega_Z\rightarrow \omega_M}(\omega_M - \omega_Z)\quadraom\\
&= \braket{\etm Y\et|l_L}\braket{\karl|\esd\etm \xmz\et\esdm  | \etrm}\braket{l_M|\etm Z\et}\nonumber.
}{prod}
It is important to notice that one obtains an approximation to the product $\TT_{0L}^Y \TT_{LM} \TT^Z_{M0}$ rather than an expression for $\TT_{LM}$.
To single out the $\TT_{LM}$ transition moment, we 
 divide the whole quantity \fr{prod} by the product of the left and right transition
moments from the ground state.
These are obtained from the  residue of
$\langle \Psi^{(1)}(\omega_Y)|X|\Psi^{(1)}(-\omega_Z) \rangle$ with $L=M$ and $Y=Z$ and $X=1$.
For the exact quadratic response function this quantity
is simply $|T_{0L}^Y|^2 = \braket{\Psi_0|Y|\Psi_L}\braket{\Psi_L|Y|\Psi_0}$, and thus can be used to extract
the transition moment between the excited states.
In the XCC theory, $|T_{0L}^Y|^2$ is a product of three integrals
\equl{|T_{0L}^Y|^2  = \braket{\etm Y\et|l_L}\braket{\kappa(r_L, S, T)|\eta(r_L, S)}\braket{l_L|\etm Y\et}. }{ppp}
As the final result, the  residue of the quadratic response
function in the XCC theory divided by $|\TT_{0L}^Y\TT_{M0}^Z| = \sqrt{|T_{0L}^Y|^2|T_{M0}^Z|^2}$ is
given by the expression
\equsl{\TT_{LM}
    & =  \pm\frac{\lim\limits_{\footnotesize{\omega_Y\rightarrow -\omega_L}}(\omega_L + \omega_Y)\lim\limits_{\omega_z\rightarrow \omega_M}(\omega_M - \omega_Z) \quadraom}
    {\sqrt{|T_{0L}^Y|^2|T_{M0}^Z|^2}}\\[10pt]
    &= 
  \pm\frac{\xi_L^Y\braket{\karl|\esd\etm \xmz\et\esdm \etrm}\xi_M^Z}
    {\sqrt{\xi_L^Y \braket{\karl|\etrl}(\xi_L^Y)^{\star}\xi_M^Z \braket{\karm|\etrm}(\xi_M^Z)^{\star}}},
}{tkn}
where $\xi_M^Z$ is defined analogously as \fr{xin}.
The sign of $\TT_{LM}^X$ is 
of concern because in practical applications one
requires only the transition strengths, i.e., the products $\TT^X_{LM}\TT^X_{ML}$.
The final expression for
$T_{LM}$ in the XCC theory is given by
\equl{\TT_{LM}
     = \frac{\braket{\karl|\esd\etm \xmz\et\esdm \etrm}}{\sqrt{\braket{\karl|\etrl}\braket{\karm|\etrm}}}.
}{glowne2}
The Hermiticity relation
\equl{\TT^X_{LM} = (\TT^X_{ML})^*
}{hermva}
is satisfied for any truncation of $T$, but does not hold exactly for a truncated  operator $S$.
 The violation of \fr{hermva} was investigated numerically
 in Ref~\citep{tucholska2017transition} and we present the results in \Frt{negat} (Table VII in Ref~\citep{tucholska2017transition}).
 For the excited-state transitions of the Mg atom
it was found  that the violation of \fr{hermva} is numerically negligible and in the problematic cases 
 does not lead to an unphysical results as in the TD-CC method.
 \begin{table}[!t]
     \begin{adjustbox}{max width=\textwidth}
   \begin{threeparttable}
     \caption{Reprinted from Ref~\citep{tucholska2017transition} with permission of AIP Publishing (Table VII, page 034108-7).
       $\TT_{LM}^X$ and $(\TT_{ML}^X)^{\star}$ computed with the TD-CC and XCC methods for the Mg atom.}
\label{negat}
\begin{tabular}{l d{2.2} d{2.2} d{2.2} d{2.2}}
    \multicolumn{1}{c}{Transition}  
    &   \multicolumn{1}{c}{$\TT_{LM}^X(\mathrm{TDCC})$} 
    &   \multicolumn{1}{c}{$(\TT_{ML}^X)^\star(\mathrm{TDCC})$}
            &   \multicolumn{1}{c}{$ \TT_{LM}^X(\mathrm{XCC})$ }
            &   \multicolumn{1}{c}{$(\TT_{ML}^X)^\star(\mathrm{XCC})$}          \\
    \hline
    \multicolumn{5}{c}{\footnotesize{aug-cc-pVQZ}\cite{prascher2011gaussian}}\\
    \hline
 $3s4s\sS - 3s3p\sP^\circ$ & 4.30 & 4.26& 4.00&4.01\\
  $3s4p\sP^\circ - 3s4s\sS$& 8.39 & 8.30 & 8.36 & 8.36 \\
     \hline
 \multicolumn{5}{c}{\footnotesize{d-aug-cc-pVQZ}\cite{feller1996role, schuchardt2007basis}}\\
            \hline
   $3s5s\sS - 3s4p\sP^\circ$ & 10.12 & 10.04 & 10.08 & 10.09 \\ 
   $3s5s\sS - 3s3p\sP^\circ$ & 0.60 & 0.60 & 0.51 & 0.51 \\
  $3s3d\sD - 3s3p\sP^\circ$ & 0.67& -0.40 & 1.40 & 1.43 \\
    $3s4p\sP - 3s3d\sD$ &-1.18 & 0.72  & 2.64 & 2.63 \\
\end{tabular}
   \end{threeparttable}
   \end{adjustbox}
\end{table}
Ref~\citep{tucholska2017transition} also reports the lifetimes of singlet and triplet states for the Ca, Sr, and Ba atoms
and presents a comparison against experimental data of the Gaussian\cite{boys1950electronic, boys1956automatic}
and Slater\cite{lesiuk2017combining} basis set implementations. The conclusions are that the use of the CC3
approximation combined with the STO basis sets
give the results that show the  smallest mean absolute
percentage deviation from experimental data. Additionally the proof that the XCC transition moments are size intensive can be found Ref.~\citep{tucholska2017transition}.

\section{Summary}
We have presented a review of the expectation value method for the computation of ground-state properties, response properties, and transition probabilities within the coupled-cluster formalism. We summarized the derivations of the average value of an observable, one- and two-electron density matrices, polarization propagator, quadratic response function, and transition moments between the ground and excited states. The XCC formalism is conceptually simple and easily extendable to general operators: the calculation of a property requires only a single-step computation of the amplitudes of the auxiliary operator S that enter the definition of the average value. The methodology can easily be extended to the CC models other CCSD and CC3 provided that the set of commutators/contributions retained in the working formulas for the matrix elements properly corresponds to the choice of the ground state amplitudes. The computational cost of the XCC computation is controlled by the level of approximation of the operator S and the subset terms retained in the commutator expansion of the property value. The selection of included contributions is guided either by the MBPT expansion or powers of T. Both approaches yield a straightforward scheme of approximations that can be expanded to include higher excitations or to employ one of modern highly-efficient computational techniques developed for the ground-state coupled cluster theory. In particular, the simple formulation of the XCC method makes it possible to combine it with the explicitly correlated methods, which is a work underway in our laboratory.  Further work related to the XCC theory concerns the description of the spin-orbit coupling matrix elements, magnetic moments, nonadiabatic coupling and open-shell systems. 
\section{Acknowledgment}
We would like to thank Dr. M. Lesiuk and Dr. M. Modrzejewski for useful discussions regarding the manuscript.
 This research was supported by the National Science Center (NCN) under Grant No. 2017/25/B/ST4/02698.

{\footnotesize \bibliography{ref}}

\end{document}